\documentclass[epj]{svjour}
\usepackage{latexsym}
\usepackage[dvips]{graphicx}
\newcommand{\be}{\begin{equation}}
\newcommand{\ee}{\end{equation}}
\newcommand{\bea}{\begin{eqnarray}}
\newcommand{\eea}{\end{eqnarray}}
 
\newcommand{\ld}{\ell_{\hbox{\tiny D}}} 
\newcommand{\jes}{j_{\hbox{\tiny ES}}} 
\newcommand{\jm}{j_{\hbox{\tiny M}}} 
\newcommand{\jsb}{j_{\hbox{\tiny SB}}} 
\newcommand{\ml}{m_{\hbox{\tiny 2L}}} 
\newcommand{\Ves}{V_{\hbox{\tiny ES}}} 
\newcommand{\Ul}{U_{\hbox{\tiny L}}} 
\newcommand{\Sm}{S_{\hbox{\tiny M}}} 
\newcommand{\egs}{\epsilon_{\hbox{\tiny GS}}} 
\newcommand{\etgs}{\tilde\epsilon_{\hbox{\tiny GS}}} 
\newcommand{\Ui}{U_{\hbox{\tiny $\infty$}}} 
\newcommand{\Ud}{U_{\hbox{\tiny 2}}} 
\newcommand{\mi}{m_{\hbox{\tiny $\infty$}}} 
\newcommand{\tmax}{t_{\hbox{\tiny MAX}}} 
\newcommand{\fra}[2]{\hbox{${#1\over #2}$}}
\begin{document}
\title{Coarsening process in one-dimensional surface growth models}
\author{Alessandro Torcini\inst{1,2,}%
\thanks{e-mail: {\tt torcini@ino.it}}
\and Paolo Politi\inst{2,}%
\thanks{Corresponding author: e-mail: {\tt politi@fi.infn.it}}
}                     
%
%
\institute{
Dipartimento di Energetica ``S. Stecco", Universit\`a di Firenze,
Via S. Marta 3, 50139 Firenze, Italy
\and 
Istituto Nazionale per la Fisica della Materia, UdR Firenze, 
Via G. Sansone 1, 50019 Sesto Fiorentino, Italy
}
\date{Received: date / Revised version: date}
%
\abstract{
Surface growth models may give rise to instabilities with mound
formation whose tipical linear size $L$ increases in time
(coarsening process). In one dimensional systems 
coarsening is generally driven by
an attractive interaction between domain walls or kinks.
This picture applies to growth models for which the largest surface slope 
remains constant in time (corresponding to model~B of dynamics):
coarsening is known to be logarithmic in the absence of noise
($L(t)\sim\ln t$) and to follow a power law ($L(t)\sim t^{1/3}$)
when noise is present.
If surface slope increases indefinitely, 
the deterministic equation looks like a
modified Cahn-Hilliard equation: here we study the 
late stages of coarsening through a linear stability analysis of
the stationary periodic configurations and through a direct 
numerical integration.
Analytical and numerical results agree with regard to the conclusion that 
steepening of mounds makes deterministic coarsening {\em faster}\,:
if $\alpha$  is the exponent describing  the steepening of the maximal slope
$M$ of mounds ($M^\alpha\sim L$) we find that
$L(t)\sim t^n$: $n$ is equal to ${1\over 4}$ for
$1\le\alpha\le 2$ and it decreases from ${1\over 4}$ to ${1\over 5}$
for $\alpha\ge 2$, according to $n=\alpha/(5\alpha -2)$. 
On the other side, the numerical solution of the corresponding stochastic
equation clearly shows that in the presence of shot noise steepening
of mounds makes coarsening 
{\em slower} than in model~B: $L(t)\sim t^{1/4}$, irrespectively of $\alpha$.
Finally, the presence of a symmetry breaking term is shown not to modify
the coarsening law of model $\alpha=1$, both in the absence and
in the presence of noise.
} 
\maketitle
\section{Introduction}
\label{intro}
\vskip 1cm
In real systems surface growth occurs on two-dimen\-sio\-nal (2d) substrates
and therefore its modelization in one dimension (1d) is in general an
oversimplification, mainly justified by the possibility to have a deeper
understanding of the dynamical evolution of the system. 
In some cases the surface indeed maintains a 1d
profile, for example when the relaxation of a grooved surface is 
studied~\cite{grooves}.
This is not true when the surface undergoes kinetic roughening 
or a growth instability followed by a phase
separation: in both cases, noise makes the resulting morphology
2d even if the initial one is 1d.
The dynamical evolution of a vicinal surface is another possible 
application of 1d models. In fact, if atomic steps move in 
phase~\cite{PRL_Saito}, step motion is described by a
one-dimensional growth equation.

In this paper we are interested in a growing surface that 
undergoes a kinetic instability.
The origin of such an instability is related to an additional barrier 
that diffusing ad\-atoms must overcome to descend step edges 
(the so-called Ehrlich-Schwoebel (ES) barrier~\cite{KE,libroJV}).
Even in the presence of stabilizing mechanisms, at sufficiently large
scales the ES effect can be the dominant one 
and the flat surface becomes unstable against small deformations. 
In a first linear regime,
a structure with a well-defined wavelength
emerges and its amplitude increases in time.
At a later stage, nonlinearities come into play and the mound structure
typically --but not unavoidably~\cite{PRL_Saito}-- coarsens. 

This simplified picture reminds spinodal decomposition and 
coarsening that take place during phase separation.
Surface growth instability and phase separation
do have strong similarities and they can indeed be equivalent in
1d~\cite{review}. However, they are definitely different in two 
dimensions~\cite{coarsening2d}.

Generally speaking, phase separation has different properties
in one and two dimensions. 
We mention here two of them~\cite{revBray}: 
(i)~In 2d, coarsening is driven by the tension of domain walls while
in 1d it is due to interaction between walls. 
(ii)~Noise is generally irrelevant in 2d, while it modifies the coarsening law
in 1d (except in the presence of long-range interactions).

The class of models that we study in one dimension is of
interest in two respects. First, since the slope is assumed to increase
indefinitely, it is not possible to speak of domain walls between
different phases. 
Second, being the interaction long ranged, noise may happen to be
irrelevant for the coarsening law in 1d:
this occurs for some of the models studied here.

In the next Section we give a short introduction 
to the continuum equations that are encountered in one-dimensional 
models of conserved surface growth. A more detailed analysis can be found 
in recent review papers~\cite{review,revJK}.        
In Section~\ref{sec:theo} we recall a few theoretical
approaches to one-dimensional coarsening and in Section~\ref{sec:det}
we apply the linear stability analysis to our class of 
deterministic $\alpha$-models. Its results are confirmed by
numerics. The problem of coarsening in the presence of noise
is addressed directly via stochastic numerical integration in
Section~\ref{sec:noise}. In Section~\ref{sec:sbt} we examine
the effect of a symmetry breaking term, while
the discussion of the results and our conclusions are 
presented in the final Section.

\section{Models for unstable surface growth}
\label{models}

In the Introduction we have already mentioned 
the ES barriers as a source of the kinetic instability.
In fact, it is well-known~\cite{VilJdP} that an
asymmetry in the sticking coefficients of an adatom to a step produces
a slope-dependent current $\jes (m)$; $h(x,t)$ and $m=\partial_x h$
are the local height and slope of the surface.
It is more convenient to use the variable $z(x,t)=h(x,t)-\bar h$,
where $\bar h = F_0 t$ is the average height, instead of $h(x,t)$
($F_0$ being the intensity of the flux). 
In this way the evolution equation $\partial_t h =F_0 -\partial_x j$ simply 
writes $\partial_t z = -\partial_x j$ while the definition of the
slope $m=\partial_x h =\partial_x z$ is unaffected.

For symmetry reasons, $\jes$ is an odd function of $m$. 
Therefore we expect that 
$\jes\simeq\nu m$ at small slopes, as confirmed by a more rigorous
analysis~\cite{review}. The asymmetry in the sticking coefficients is
generally due to the additional energy barrier hindering the adatom
from descending a step. This gives rise to an up-hill current
$\jes\simeq\nu m$ with a positive coefficient $\nu$.
This implies that $\jes$ has a destabilizing character, as easily revealed by
the solution of the linear differential equation
$\partial_t z = -\partial_x \jes = -\nu \partial_x^2 z(x,t)$~:
$z(x,t)=z_0\cos(qx)e^{\omega_q t}$, with $\omega_q = \nu q^2$.

In a continuum model, where the lattice constant $a$ goes to zero,
the following expression for $\jes$ can be used:
\be
\jes = {\nu m\over 1+\ld^2 m^2} ~~~~~~\hbox{model 1} ~ .
\label{mod1}
\ee

If a discrete model with square symmetry is used,
we expect $\jes$ to vanish for $m=m_0=1/a$ . We can therefore define 
the model:
\be
\jes = \nu m(1-m^2/m_0^2) ~~~\hbox{model 0} 
\label{mod0} 
\ee

However, other mechanisms can produce a slope-de\-pen\-dent current:
short-range step-adatom interaction~\cite{KE,AF}, post-deposition
transient mobility and downward funneling~\cite{Evans}.
The first mechanism may be either stabilizing or not, 
depending on the sign of
the step-adatom interaction while the other (non-thermal) mechanisms
are typically stabilizing, i.e. they contribute with a negative
term $-\nu' m$ to $\jes$.
Because of that, either the ES current acquires a
zero at finite slope (from model~1 we pass to model~0) 
or there is a change in the value $m_0$ at which $\jes$ vanishes.

Finally, in a previous paper~\cite{JPA} we have generalized 
Eq.~(\ref{mod1}) to
a class of models ($\alpha$-models) characterized by different 
asymptotic behaviors for $m\to\infty$:
\be
\jes = {\nu m\over (1+\ld^2 m^2)^\alpha} ~~~~~~
\hbox{model }\alpha~(\ge 1) ~ . \label{modalpha}
\ee

In the Introduction we also mentioned 
possible stabilizing mechanisms. The simplest expression for a
stabilizing current is the so-called Mullins term that in its
linearized form reads $\jm=K\partial_x^2 m$. Its origin may be either
thermodynamic (relaxation through surface diffusion~\cite{Mullins})
or kinetic (fluctuations in the nucleation process of new 
islands~\cite{PV,Nato_Rodi}).
A stabilizing current gives a negative contribution
to $\omega_q$. Starting from the
equation $\partial_t z = -\partial_x (\jes + \jm)$ it is easily found
that $\omega_q = \nu q^2 - K q^4$. A flat surface
is thereby linearly unstable against fluctuations of wavelength larger
than $\lambda_c = 2\pi\sqrt{K/\nu}$.

Both currents $\jes$ and $\jm$ change sign if
$x\to -x$ or $z\to -z$: the growth process cannot break the former symmetry
but it does break the latter.
It has been shown~\cite{VilJdP,PV} that such symmetry breaking term 
(intrinsically nonlinear)
has the form $\jsb=\partial_x A(m^2)$, where $A\sim m^2$ at small slopes and
$A\sim -1/m^2$ at large ones. Its presence is strictly related to the
breaking of the detailed balance principle~\cite{Racz} and therefore to
the non-equilibrium character of the growth process.

It has been proven~\cite{kinks} that $\jsb$ does not modify the coarsening law
of model~0: the effects of $\jsb$ on 
model~1 will be considered in Section~\ref{sec:sbt}.

We conclude this Section 
by writing down explicitly the class of growth equations
that are analyzed in the paper:
\be
\partial_t z(x,t) = - \partial_x j(x,t) + \eta(x,t)\label{1}
\label{model}
\ee
\be
j(x,t) = \left\{ 
\begin{array}{lr}
\partial_x^2 m + {m\over (1+m^2)^\alpha} & \hbox{model }\alpha \cr
\partial_x^2 m + m(1-m^2) & \hbox{model }0 
\end{array}  \label{2} \right.
\ee
\be
\langle\eta(x,t)\rangle =  0 ~~~ 
\langle\eta(x,t)\eta(x',t')\rangle = \tilde F_0 \delta(x-x')\delta(t-t')
\label{4}
\ee

Eq.~(\ref{1}) is the evolution equation of the local height $z(x,t)$ for
a conserved growth process in the presence of the shot noise $\eta(x,t)$;
Eqs.~(\ref{2}) give the currents for the $\alpha$-models
and model~0, once that
$x,t,z$ have been rescaled in order to get an adimensional equation;
Eq.~(\ref{4}) gives the spectral properties of the noise, whose strength
$\tilde F_0 = F_0a\ell^2/\sqrt{\nu K}$ is the only parameter 
appearing in the problem:
$F_0$ is the intensity of the flux, $a$ is the lattice constant,
$\ell$ is the diffusione length $\ld$ for $\alpha$-models 
(Eqs.~(\ref{mod1},\ref{modalpha})) or the inverse of the constant slope 
$m_0$ for model~0 (Eq.~(\ref{mod0})), $\nu$ and $K$ are the
prefactors of $\jes$ and $\jm$, respectively.

\section{Theoretical approaches to coarsening}
\label{sec:theo}

In this Section we review two theoretical methods that have
been used to study the coarsening process in model~0.
The first method uses the property that the current $\jes$ vanishes at a finite
slope $m_0$. 
Although it can not be used for $\alpha$-models, for 
completeness and clarity we discuss it in Section~\ref{ss:kinks}.
The second method consists in a linear stability analysis of the
stationary configurations. It is a very general method and it is
discussed in Section~\ref{sec:lsa}.
In the context of phase separation processes it was introduced by 
Langer~\cite{Langer} to study model~0 and its solution is proposed in 
Section~\ref{sec:mod0}. Its application to $\alpha$-models is carried out
in Section~\ref{ss:alpha_models}.

\subsection{Kink dynamics}
\label{ss:kinks}

Model 0 and $\alpha$-models have the same linear behaviour, but 
they strongly differ in the nonlinear regime: in model~0 the slope
increases up to the maximal value $m_0=\pm1$ while in $\alpha$-models
it grows indefinitely. 

For model 0, the surface profile corresponding to a constant value $m_0=1$
(i.e. to a vicinal surface with unitary slope) is stable~\cite{note2}, 
but it can not be
attained starting from a flat surface because the average slope must remain
constant. Still, we can consider the stationary configuration $m_+(x)$
($j(m_+(x))\equiv 0$) corresponding to a limiting slope $\pm m_0$ 
for $x\to\mp\infty$. That profile is called `mound'
in the $z$-space and `kink' or `domain wall' in the space of the order
parameter $m$, where it has the form $m_+(x)=\tanh(x/\sqrt{2})$. 

During the coarsening process of model~0 a typical surface profile is
just an alternating sequence of kinks ($m=m_+(x)$) and antikinks
($m=m_-(x)=-m_+(x)$), whose average distance $L(t)$ increases
with time because of the annihilation process between pairs
of neighbouring kink-antikink. In the absence of noise the dynamics of kinks is
governed by their attractive interaction 
that decays exponentially with the distance, since
$|m_\pm (x)| \simeq 1 - 2\exp(-\sqrt{2}|x|)$ for $|x|\gg 1$.
Such a weak interaction determines a very slow coarsening:
$L(t)\sim \ln t$~\cite{giap,Langer}. 

In the presence of noise the picture is different because of the induced
fluctuations on the kink positions.
If there were no constraint induced by the conservation of the 
order parameter, kinks would simply perform
a random walk and therefore would travel a distance $L$ in a
typical time $L^2$, giving a coarsening exponent $n=1/2$ 
($L(t)\sim\sqrt{t}$).
In a growth process, where a conservation law does exist, 
kink trajectories are not independent and the coarsening slows down: $n=1/3$.
This exponent is more easily understood in a spin picture~\cite{CKS},
where the conservation of the order parameter (the magnetization) implies that
the system evolves through spin-exchange processes (Kawasaki dynamics).

If we now turn to $\alpha$-models, we recognize that the kink picture
is not applicable because the unstable current $\jes$ vanishes for
infinite slope only. 

\subsection{Linear stability analysis of stationary configurations}
\label{sec:lsa}

We now discuss the linear stability
analysis of the stationary configurations. 
They are determined by
the condition $\partial_t z(x,t)\equiv 0$, i.e. the current
must be a constant: $j\equiv c$. 
The net current $c$ is related to the average slope of the surface: 
we are interested in a high symmetry surface and therefore 
we set $c=0$.

The equation $j=0$ is formally equal to
the Newton's equation for a particle of unitary mass, where the slope 
$m$ plays the role of the particle position and $x$ the role of time:
\be
\partial_x^2 m(x) + \jes(m) = 0 ~ .
\label{j=0}
\ee

\begin{figure}
\begin{center}
\includegraphics[angle=-90,width=8.5cm,clip]{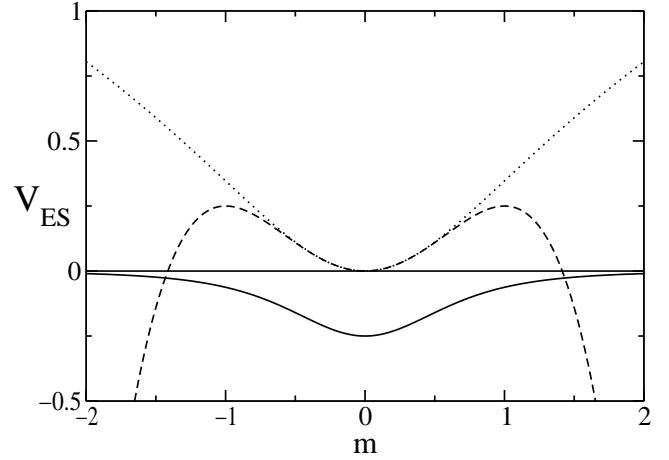}      
\end{center}
\caption{Profiles of the different potentials $\Ves$. From top
to bottom: Model~1 (dotted line), Eq.~(\protect\ref{U1}); 
Model~0 (dashed line), Eq.~(\protect\ref{U0}); 
Model~$\alpha=3$ (full line), Eq.~(\protect\ref{U2}).}
\label{fig:potenziali}
\end{figure}

The fictitious particle feels the force $-\jes(m)$, i.e. it moves in
the potential $\Ves(m) = \int^m ds \jes(s)$. Different models give
qualitatively different potentials (see Fig.~\ref{fig:potenziali}):
\bea
\Ves(m) &=& m^2/2 - m^4/4  ~~~~~~~~~~~~~\hbox{model 0} \label{U0}\\
\Ves(m) &=& (1/2)\ln(1+m^2)   ~~~~~~~~~~\hbox{model 1} \label{U1}\\
\Ves(m) &=& -(1+m^2)^{1-\alpha}/[2(\alpha -1)] ~
\hbox{model }\alpha>1 \label{U2}
\eea

Stationary configurations therefore correspond to the periodic
oscillations of the particle around the minimum of $\Ves$ in $m=0$.
We label the stationary configurations $\ml(x)$
through their period $2L$. What about the limit $L\to\infty\;$?
For model~0 it corresponds to the kink-solution: 
$\mi(x)=m_+(x)\to \pm 1$ when $x\to\pm\infty$. For model~1, the energy of the
particle diverges when the period $L$ increases and the limiting
configuration $\mi(x)$ does {\it not} exist.
It does exist for $\alpha>1$ and it corresponds to the
well defined problem of a particle of zero energy moving in the 
potential (\ref{U2}): it starts at $m=-\infty$ and arrives 
at $m=+\infty$ after an infinite time.

We start by considering small deviations $\psi$
from the periodic profile: $m(x,t)=\ml(x) + \psi(x,t)$.
Since $\jes(\ml + \psi)= \jes(\ml) + \jes'(\ml)\psi + {\cal O}(\psi^2)$, 
we obtain that the linearized evolution equation for $\psi(x,t)$ is
\be
\partial_t \psi = \partial_x^2\left[ -\psi''(x,t) -\jes'(\ml(x))\psi\right]
\label{eq8}
\ee   
and therefore the time dependence of $\psi$ is: 
$\psi(x,t)=\phi(x)\cdot$ $\exp(-\epsilon t)$. 
Replacing the expression of $\psi$ in terms of $\phi$ into Eq.~(\ref{eq8}), 
we find that the stability is determined by the spectrum of the following 
operator:
\be
(-\partial_x^2)\left[ -\phi''(x) + \Ul (x)\phi\right]
\equiv D_x\hat H\phi (x) =\epsilon\phi~,
\label{eq9}
\ee 
where $D_x\equiv -\partial_x^2$ and $\hat H\equiv -\partial_x^2 + \Ul(x)$
is a single-particle Schroedinger operator corresponding to the
periodic potential $\Ul (x) \equiv -\jes'(\ml(x))$, 
{\it of period} $L$:
 
\bea
\Ul(x+L) &=& -\jes'(\ml(x+L)) \nonumber\\ 
&=& -\jes'(-\ml(x)) = \Ul(x) ~ .
\eea

In one dimension coarsening is due to the unstable character
of the periodic stationary configurations, i.e. to the existence of
negative eigenvalues in the energy spectrum~\cite{nota1}. 
Because of the periodic
character of the operator $D_x\hat H$, eigenvalues are grouped into 
bands. 

Our evolution equation for $m(x,t)$ is $\partial_t m=D_x j$ where the
current $j$ (see Eq.~(\ref{2})) can be derived from a pseudo free 
energy ${\cal F}$:
\be
j= - {\delta {\cal F}\over \delta m} ~~~~~~~
{\cal F} = \int dx [\fra{1}{2} (\partial_x m)^2  - \Ves(m)]
\ee

For model~0 the potential $-\Ves(m)$ has the standard double well shape 
and $m(x,t)$ evolves accordingly to the Cahn-Hilliard equation;
in the presence of conserved noise we obtain the so-called model~B of 
dynamics~\cite{HH}. If $D_x$ is replaced by the identity
operator, the order parameter
is no longer conserved and its evolution equation is 
$\partial_t m=j$, which is equivalent for model~0 to the time 
dependent Ginzburg
Landau equation, or --in the presence of nonconserved noise-- to 
model~A of dynamics~\cite{HH}.
We use the notations $\tilde\epsilon,\tilde\phi$ for the
spectrum of the operator $D_x\hat H$ and 
$\epsilon,\phi$ for the Hamiltonian operator $\hat H$.

Let us start with few general statements on the lowest part of the spectrum. 
Translational invariance implies that $\epsilon=0$ is
always an eigenvalue of the operator $\hat H$ (and therefore
$\tilde\epsilon=0$ is an eigenvalue of
$D_x\hat H$ as well) and it corresponds to the eigenfunction
$\phi(x)=\tilde\phi(x)=\ml'(x)$. To prove it let us use the definition
of $\ml(x)$ as solution of the differential equation~(\ref{j=0}) and take
its derivative:
\be
\ml'''(x) + \jes'(\ml)\ml'(x) = 0~.
\ee
Since $\Ul(x)=-\jes'(\ml(x))$ we just have
\be
-\partial^2_x (\ml'(x)) + \Ul(x)(\ml'(x)) \equiv \hat H \ml'(x) = 0~.
\ee

Therefore the operators $\hat H$ and $D_x\hat H$ have a zero energy
eigenvalue and the corresponding eigenfunction $\ml'(x)$ has
period $\lambda=2L$ (i.e. it corresponds to  the wavevector
$q=\pi/L$ in the Bloch representation). 
 
We can now recognize the importance of the limit
$\mi(x)$. If $\mi(x)$ exists, for $L\to\infty$ 
the periodic potential
$\Ul(x)$ becomes a single well potential $\Ui(x)$. 
In this limit $\ml(x)$ is a monotonic function
$m_+(x)$ (the kink solution, for model~0). 
Therefore $\tilde\phi_1(x)=\phi_1(x)\equiv m_+'(x)$ 
has no node and it represents
the ground state for the single well problem. For finite $L$ we have a
periodic potential and the energy level $\epsilon_1=\tilde\epsilon_1=0$ 
gives rise to the lowest band of the spectrum. 
The ground state $\egs(L)$ of the operator $\hat H$ corresponds
to $q=0$ and must therefore have a negative energy, implying that
$\hat H$ has negative eigenvalues.
The relation $\etgs=\tilde\epsilon(q=0)$
is no longer valid for the conserved model, but Langer~\cite{Langer}
has shown that negative eigenvalues $\epsilon$ of $\hat H$ 
may be put in correspondance to negative eigenvalues $\tilde\epsilon$
of $D_x\hat H$ (see Section~\ref{sec:mod0}).

Since an unstable mode increases exponentially with the
factor $\exp(|\epsilon |t)$, the knowledge of the $L$ dependence
of the ground state energy allows to find the deterministic
coarsening law $L(t)$ {\it via} the relations $|\egs(L)|\sim 1/t$
(nonconserved model) and $|\etgs(L)|\sim 1/t$ (conserved model).

\section{Deterministic coarsening}
\label{sec:det}

\begin{figure}
\begin{center}
\includegraphics[angle=-90,width=8.5cm,clip]{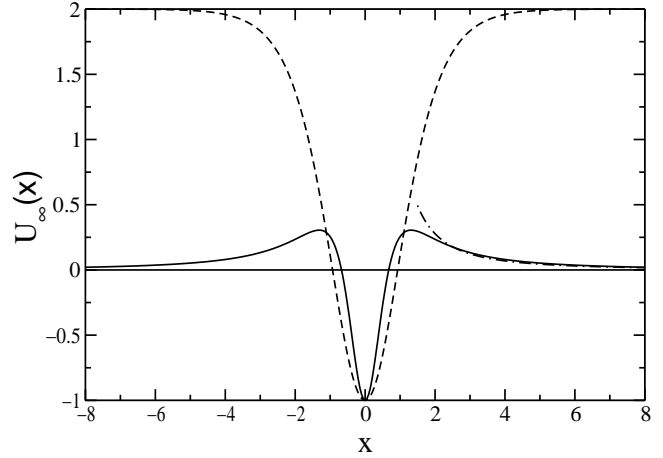}      
\end{center}
\caption{The single well potential $\Ui(x)$ for model~0
(dashed line) and for model $\alpha=3$ (full line).
At large $x$ the former approaches 2,
while the latter vanishes as 
$\Ui(x)=a/x^2$, with $a=10/9$ (dashed-dotted line, 
see formula~(\ref{Udix_asi})). }
\label{fig:singola_buca}
\end{figure}

In Fig.~\ref{fig:singola_buca} we plot the single well potentials
$\Ui(x)$ for model~0 (dashed line) and for model $\alpha=3$ (full line),
the latter being representative of all the class of $\alpha$-models.
Since, as already explained, $\epsilon_1=0$ is the ground  state energy
for the single well, completely different behaviours
are expected in the two cases. 

For model~0 (see Section~\ref{sec:mod0}), 
$\Ui(x)$ approaches 2 at large $x$:
therefore the wavefunction $\phi_1(x)$ 
decays exponentially at large distances and energy shifts due to 
the tunneling between wells at finite distance $L$ are expected to
be exponentially small.
This property is the counterpart of the exponentially
vanishing interaction between kinks, discussed in Section~\ref{ss:kinks}.

For $\alpha$-models (see Section~\ref{ss:alpha_models}), 
$\Ui(|x|=\infty)=0=\epsilon_1$
and the considerations developed for model~0 do not apply. 

\subsection{Model 0}
\label{sec:mod0}

For model~0 the kink-solution is $\mi(x) = \tanh (x/\sqrt{2})$, the
single well potential is (see the dashed line in Fig.~\ref{fig:singola_buca})
\be
\Ui(x) = -1 + 3\mi^2(x) = -1 + 3\tanh^2 (x/\sqrt{2})
\ee
and the corresponding ground state wavefunction is 
$\phi_1(x)=\mi'(x)=\fra{1}{\sqrt{2}}\sec^2(\fra{x}{\sqrt{2}})$.

For finite $L$, $\Ul (x)$ is a collection of
wells centred at points $x=0,\pm L, \pm 2L, \dots$~.
It is interesting to compare $\Ul(x)$ with the periodic potential
$\Ul^*(x)$ obtained as a superposition of the single well potentials
$\Ui(x-nL)$:
\be
\Ul^*(x) = \Ui(x) + \sum_{n\ne 0} [ \Ui(x-nL) - \Ui(\infty) ] ~ .
\label{ulstar}
\ee

\begin{figure}
\begin{center}
\includegraphics[angle=-90,width=8.5cm,clip]{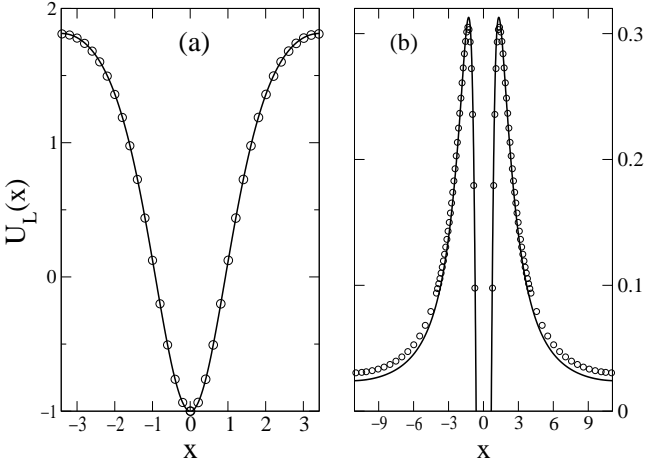}      
\end{center}
\caption{The true periodic potential $\Ul(x)$ (empty circles)
is compared with the potential $\Ul^*(x)$ (full line), obtained as a
superposition of the single well potentials $\Ui(x-nL)$
(see Eq.~(\protect\ref{ulstar})).
We make this comparison for model~0~(a), where $L\simeq 6.8$, 
and for model~$\alpha=$3~(b), where $L\simeq 22.8$.
In both cases we display one period only.
$\Ul(x)$ is obtained numerically and $\Ul^*(x)$ is obtained
analitically. The superposition principle works well in (a)
even for small $L$, while the
exact potential for model~3 (see b) is {\it not} reproduced by 
the sum of single well potentials.
In order to make more evident the difference between the two potentials,
in (b) we focus the plot on the region of positive $\Ul(x)$.
We remind that $\Ul(0)=-1$ for any $\alpha$.}
\label{fig:somma_buche}
\end{figure}

For model~0, $\Ui(\infty)=2$ and the summation can indeed be limited to the
two terms $n=\pm 1$ because the quantity in square brackets
is exponentially small.
In Fig.~\ref{fig:somma_buche}a 
we show that $\Ul^*(x)$ (full line) is an excellent approximation
of $\Ul(x)$ (circles).

Langer performed a tight-binding-approximation to 
determine the lowest band of the energy spectrum arising from the
ground state $\epsilon_1=0$ of the single well.
The result~\cite{Langer} is
\be
\epsilon(q) \simeq  -(1+\cos qL)\exp(-2L)~,
\label{eq_l:6.15}
\ee
confirming that $\epsilon(q=\pi/L)=0$, $\egs=\epsilon(q=0)$
and that $\egs$ decays exponentially with the
distance $L$ between wells. The relation $|\egs|\sim 1/t$~\cite{nota2}
implies that coarsening is logarithmically slow:
$L(t)\sim \ln t$. This result is valid for the {\it conserved}
model as well. Langer proved it by means of the variational
condition~\cite{Langer}:
\be
\etgs \le {\egs\over (\bar\phi_1,\phi_1) }
\ee
where 
\be
(\bar\phi_1,\phi_1)=\int dx\bar\phi_1^*(x)\phi_1(x)
\ee
is the scalar product between the ground state function $\phi_1(x)$
of the Hamiltonian operator and $\bar\phi_1$ is defined by the
relation $D_x\bar\phi_1=\phi_1$. 

Because of the conservation law, the lowest energy band for the
operator $D_x\hat H$ reads~\cite{Langer}:
\be
\epsilon(q) \simeq  -\sin^2 qL\cdot{\exp(-2L)\over L} ~ .
\label{eq_l:6.21}
\ee

The factor $L$ at denominator is irrelevant and the
coarsening law $L(t)\sim \ln t$ is still valid.
Eq.~(\ref{eq_l:6.21}) also confirms that $\tilde\epsilon(q=\pi/L)=0$
and that $\tilde\epsilon(q=0)$ is no more the ground state.

\subsection{$\alpha$-models}
\label{ss:alpha_models}

The solution of model~0 by Langer is of great interest because
it gives approximate expressions for the lowest energy band, 
both for the nonconserved~(\ref{eq_l:6.15}) and 
conserved~(\ref{eq_l:6.21}) models.
The basis of his treatment is the tight-binding-approximation:
$\Delta U(x)\equiv \Ul^*(x) - \Ui(x)$ is taken as a small perturbation
of $\Ui(x)$.

\begin{figure}
\begin{center}
\includegraphics[angle=-90,width=8.5cm,clip]{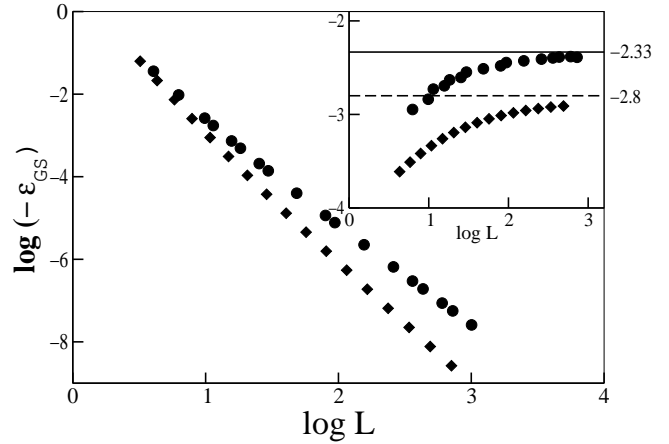}      
\end{center}
\caption{Nonconserved models: on a log-log scale (main) we report the absolute
value of the ground state energy $\egs(L)$ for the periodic potential
$\Ul(x)$. Circles refer to model~3 and diamonds to model~10.
In the inset we report the local derivatives 
$d[\log(-\egs)]/d[\log(L)]$ and the asymptotic values 
of $\gamma$ that
are calculated analytically (see Eq.~(\protect\ref{exp_gamma})).}
\label{fig:shift_noncon}                                                   
\end{figure}

In the case of $\alpha$-models, a simpler strategy can be 
followed~\cite{JPA} by replacing the periodic
potential $\Ul(x)$ with a double well $\Ud(x)$,
obtained by joining rather
than superposing $\Ui(x)$ and $\Ui(x-L)$~\cite{nota3}.
Using this approximation, our analytical results for the
coarsening exponent $n(\alpha)$ agre well with the estimate obtained
from numerical integration of the growth equations~\cite{JPA}
(see the results for the conserved model reported in
Fig.~\ref{fig:n_alpha}).
In the following we report the general lines of this
theoretical approach, and we add a direct numerical 
confirmation (see Fig.~\ref{fig:shift_noncon}).

The single well potential $\Ui(x)$ for the $\alpha$-models 
(see Fig.~\ref{fig:singola_buca})
decays to zero at large $x$ from positive values.
 From the relation $\Ui(x)=-\jes'(\mi(x))$ it is simple to derive that 
for $m\gg 1$:
\be
\Ui(x) \sim {(2\alpha -1)\over \mi^{2\alpha}(x) } ~ .
\ee

The integration of Newton's equation~(\ref{j=0}) for zero energy 
gives the asymptotic result ($x\gg 1$):
\be
\mi^\alpha(x)\sim \fra{\alpha}{\sqrt{\alpha -1}}x~,
\ee
which implies
\be
\Ui(x) \sim {(2\alpha -1)(\alpha -1)\over \alpha^2 } {1\over x^2}
\equiv {a\over x^2 }~.
\label{Udix_asi}
\ee

Thus the single well potential decays as the inverse of a square law
{\it whatever} is $\alpha$, but with a prefactor $a$ that increases
between $a=0$ for $\alpha=1$ and $a=2$, for $\alpha=\infty$.
For $\alpha=3$, $a=\fra{10}{9}$ and the corresponding 
function $\fra{a}{x^2}$ is
reported in Fig.~\ref{fig:singola_buca} as the dashed-dotted line,
showing the comparison between the asymptotic expansion (\ref{Udix_asi})
and the exact expression~\cite{nota4}.

The solution $\phi_1(x)$ of the Schroedinger equation for the single well 
(let us remind that the ground state has zero energy) therefore decays
at large $x$ as a power law~\cite{JPA}: $\phi_1(x)\sim |x|^{-\beta}$, with
$\beta=(1-\alpha^{-1})$. 
So, the ground state wavefunction is a bound state
for $\beta>\fra{1}{2}$ i.e. $\alpha>2$ only. 

In Ref.~\cite{JPA} we used the Landau and Lifshitz approach \cite{LL}
for the double-well problem and we extended it to take into
account the possibility that the ground state wavefunction for the
single well is not a bound state. This happens for $1<\alpha\le 2$.
The main point is that even if $\phi_1(x)$ is not a bound state, the ground
state $\phi_2(x)$ of the double well {\it is} bound, because its
energy $\epsilon_2$ is now strictly negative and so lower than the asymptotic
value $U_2(\infty)=0$.

Following the above approach we found~\cite{JPA} that 
\be
|\egs(L)| \simeq L^{-\gamma} ~~~~~
\gamma = \left\{
\begin{tabular}{cc}
2 & $\alpha<2$  \cr
($3-\fra{2}{\alpha}$) & $\alpha > 2$
\end{tabular}
\right.  ~ .
\label{exp_gamma}
\ee

Once $\gamma$ is known, the coarsening exponent $n$ 
is just given by $n=1/\gamma$. In Table~\ref{tab:n} we summarize 
the results found in Ref.~\cite{JPA} for the
nonconserved and conserved models.
For $\alpha=2$ there are logarithmic corrections: $L\sim (t/\ln t)^n$.

\begin{table}
\caption{The deterministic coarsening exponent $n$, for the
nonconserved ($D_x\equiv 1$) and the conserved 
($D_x\equiv -\partial_x^2$) models. For $\alpha=2$ there are
logarithmic corrections (see~\protect\cite{JPA}).}
\label{tab:n}      
\begin{center}
\begin{tabular}{c|c|c}
\hline\noalign{\smallskip}
\phantom{nonconserved} & $1<\alpha < 2$ & $\alpha >2$  \\
\noalign{\smallskip}\hline\noalign{\smallskip}
nonconserved & $\fra{1}{2}$ & ${\alpha\over 3\alpha -2}$ \\
  &  & \\
conserved & $\fra{1}{4}$ & ${\alpha\over 5\alpha -2}$ \\  
\noalign{\smallskip}\hline
\end{tabular}
\end{center}
\end{table}

The reason why it has not been possible to treat the periodic potential
$\Ul(x)$ in a more rigorous way is given in Fig.~\ref{fig:somma_buche}b,
where it is clearly shown that the superposition
principle does {\it not} work for $\alpha$-models: accordingly,
the potential $\Ul(x)$ can not be approximated as the sum of single-well
potentials. Therefore, the application of the 
tight-binding-approximation is not straightforward be\-cau\-se we do not
know the explicit expression of the perturbation
$\Delta U(x) \equiv \Ul(x) - U_\infty (x)$.
The disagreement between $\Ul(x)$ and $\Ul^*(x)$ might appear at first sight
to be unimportant, but if we neglect their difference we obtain wrong results
(as we have verified).

However, it is possible to check {\it in a direct way} 
the accuracy of our theory:
we solve numerically the quantum mechanical problem to
determine the ground state energy $\egs(L)$ of the full periodic 
potential $\Ul(x)$ (see Fig.~\ref{fig:shift_noncon} and 
footnote~\cite{nota_fig}). 
The numerical results for the exponent $\gamma$ agree fairly well, 
at large $L$, with the theoretical predictions $\gamma(3)=2.\bar 3$
and $\gamma(10)=2.8$.
This confirms that our theoretical approach to calculate 
$\egs(L)$ is indeed correct.

\subsection{Numerical analysis}

Let us now discuss our numerical results for the deterministic 
$\alpha$-models.
We have numerically integrated the equation of motion
(\ref{model}) with $\eta=0$,
starting from an initial profile $z(x,0)=r_x$, where $r_x$ is
a random variable with a flat distribution in the interval
$[-0.1,0.1]$. 

Tipically, we have followed the dynamical evolution
for a total time $\tmax \sim 400,000 - 1,600,000$, for a 
chain length $L_c=1024$, with a spatial resolution
$\delta x = 0.25$ and an integration time step $\tau=0.05$. 
A few tests have been also performed with a smaller time step
$\tau = 0.025$ and with longer chains ($L_c=2048-4096$),
obtaining consistent results. The adopted integration scheme
is a time-splitting pseudo-spectral
code: more details are reported in App.~\ref{app:det}.

\begin{figure}
\begin{center}
\includegraphics[angle=-90,width=8.5cm,clip]{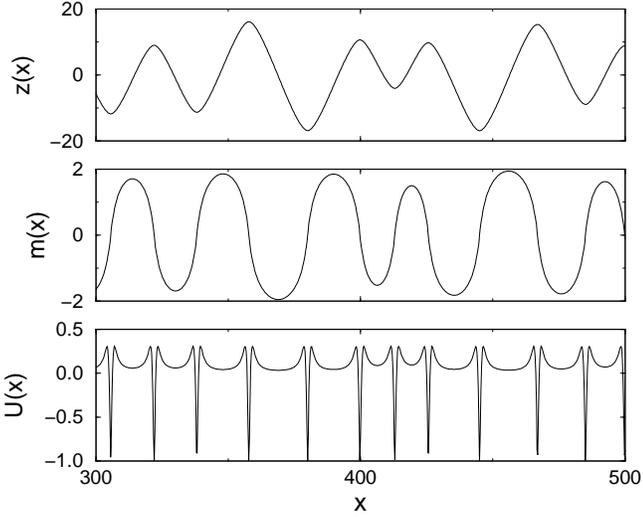}      
\end{center}
\caption{Model $\alpha=3$, late stages of coarsening.
Top: the surface profile $z(x)$ as obtained via numerical
integration. Centre: the slope profile $m(x)=z'(x)$.
Bottom: the potential $U(x)=-\jes'(m)$.
We display only a piece of the total spatial domain.
}
\label{fig:profili}
\end{figure}

On top of Fig.~\ref{fig:profili} we display a portion of the surface
profile $z(x)$ for the model $\alpha=3$. It appears 
as made up of constant slope regions separated by 
domain walls.
However, the slope profile $m(x)$ reported in the centre of
Fig.~\ref{fig:profili} does not corroborate this picture:
no region of constant slope is clearly visible and the maxima or minima
have appreciably different values.
The same remark is applicable at later times. On the bottom of
Fig.~\ref{fig:profili} we also display the potential
$U(x)=$ $-\jes'(m(x))$ that enters in the analytical solution
of the problem (see the previous section).
It is therefore reassuring that $U(x)$ looks 
indeed as a regular sequence of the single well potentials depicted in
Fig.~\ref{fig:singola_buca} as a full line,
because it confirms that the surface profile keeps close to
a stationary configuration.

The next step is to evaluate the characteristic length $L(t)$,
corresponding to the average distance between wells.
We define the wavevector $p_c(t)$ via the relation:
\be
p_c(t) = \frac{\sum_p^\prime p S(p,t)}{\sum_p^\prime S(p,t)} ~ ,
\ee
where $S(p,t)= |{\tilde z}(p,t)|^2$ is the power spectrum 
associated to the field $z$ at time $t$ 
($\tilde z$ being its spatial Fourier transform)
and the sum is restricted
to the wavevectors $p$ for which $ S(p) \ge \delta\cdot \Sm$,
$\Sm$ being the maximum value of the spectrum and
$\delta$ some threshold (typical values are  $\delta \sim 0.1 - 0.2$).
The characteristic length is then evaluated as
$L(t) = 2 \pi / p_c(t)$ and the coarsening exponent
$n(\alpha)$ has been obtained by considering the
scaling behaviour of $L(t) \sim t^n$  
in a time interval $ 10,000 < t < 400,000 - 1,600,000$.

As an independent check 
we have also determined $L$ from the normalized
spatial correlation function of the surface profile
\be
C(r,t) = \frac{<z(x+r,t) z(x,t)> - <z(x,t)>^2}{<z(x,t)^2>-<z(x,t)>^2}~,
\label{eq:funz_corr}
\ee
where the spatial average $< \cdot >$ is performed along the chain.
Defining $L$ through the relation
\be
C(L,t) = C(0,t)/2
\ee
our results are in agreement with
the previous ones obtained from the power spectrum
in all the considered cases.

\begin{figure}
\begin{center}
\includegraphics[angle=-90,width=8.5cm,clip]{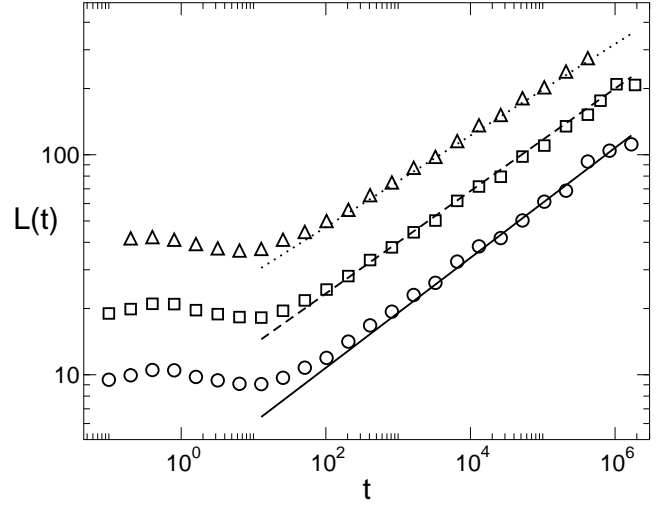}      
\end{center}
\caption{Deterministic coarsening: $L$ {\em vs} $t$ in log-log scale for 
$\alpha=1.5$ ($\circ$), $\alpha=3$ ($\Box$) and $\alpha=10$ ($\triangle$).
For presentation purpose 
the data for $\alpha=3$ and $10$ are shifted by a constant.
The lines indicate the best fit to the data
for $ t > 10,000$ and the slopes are equal to
0.250 (\hbox{---}), 0.233 (- - -) and 0.208 ($\cdots$),
in agreement with the theoretical prediction 
(see Table~\protect\ref{tab:n}). }
\label{fig:L_det}
\end{figure}

The numerically estimated length $L(t)$ is reported in 
Fig. \ref{fig:L_det} for $\alpha= 1.5,3,10$.
The coarsening exponents are 
$n(1.5)=0.250 \pm 0.003$, $n(3)=0.233\pm 0.005$ and
$n(10)=0.208 \pm 0.002$. These results are consistent
with the theoretical estimates for $n(\alpha)$, 
as summarized in Fig.~\ref{fig:n_alpha}.
Their critical discussion is deferred to the final section.

\section{Coarsening with conservative noise}
\label{sec:noise}

The stochastic equation~(\ref{model}) has been integrated up to
a time $\tmax \sim 800,000$ with $L_c=1024$. The results
for $L(t)$ and $C(r,t)$
have been obtained by averaging over $N_c$
different initial conditions with different 
noise realizations (tipically $N_c = 10$).
The integration scheme
employed in the noisy case is different from the 
one adopted for the deterministic case 
and it is described in detail in App.~\ref{app:noise}.
We have used a noise strength corresponding to a value of
$\tilde F_0$ (see Eq.~(\ref{4})) equal to 0.05.
This is a physically reasonable value because for
large ES barriers~\cite{review}, $\tilde F_0\approx\fra{a}{\ld}$
and $\ld$ is typically of the order of a few dozens of
lattice constants.

For the noisy case we have defined $L$
only in term of the average correlation function 
$\overline{C(r,t)}$, 
where the bar means that average is now
performed at each time not only along the chain (see definition
~(\ref{eq:funz_corr})) but also over 
different noise realizations. A sufficiently good scaling is obtained
in the time interval $40,000 < t < 800,000$ for all the
considered values of $\alpha$ ($=1,2,10$).

\begin{figure}
\begin{center}
\includegraphics[angle=-90,width=8.5cm,clip]{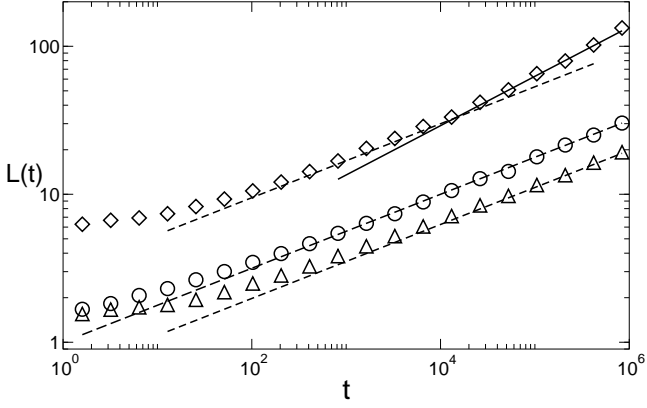}      
\end{center}
\caption{Coarsening in the presence of shot noise: $L$ {\em vs} $t$ 
in log-log scale for model~0 ($\Diamond$) and for $\alpha=1$
($\triangle$) and $\alpha=10$ ($\circ$).
Dashed lines have slope $1/4$ and full line has slope $1/3$. }
\label{fig:L_noise}
\end{figure}

As a benchmark to verify the validity both of our integration scheme
and of our procedure to estimate $n(\alpha)$, we have
analyzed model~0. In this case the coarsening exponent 
is known~\cite{giap,Langer} to be $n=1/3$. A good 
agreement between our numerical data and the theoretical
prediction is found for $ t > 40,000$, as shown in Fig. \ref{fig:L_noise}.

For the two values of $\alpha$, $\alpha=1$ and $\alpha=10$,
we find $n=1/4$. We conclude that,
in the presence of shot noise, the coarsening exponent is independent of $\alpha$
and equal to $1/4$. 
Fig.~\ref{fig:L_noise} also suggests the possible existence for model~0
of an intermediate regime (with $L\sim 5-20$) where an effective
exponent $n=\fra{1}{4}$ is found.

\section{Effects of a symmetry breaking term}
\label{sec:sbt}

In Ref.~\cite{kinks} one of us studied the effect of symmetry
breaking on model~0. Since in that model the slope keeps finite with a maximal
value equal to one, the detailed expression of the function
$A(m^2)$ ($\jsb=\partial_x A$) is not relevant.
It was therefore chosen the simplest form, the one valid
at small slopes: $A(m^2) = \lambda^* m^2$.

On the contrary, for $\alpha$-models the slope can diverge so that
the exact expression of $\jsb$ should be used~\cite{PV}:
\be
\jsb = - \lambda^* \partial_x \left( {1\over 1+m^2} \right) ~ .
\ee

We limited ourselves to the physically relevant case 
$\alpha=1$. We have therefore integrated the following differential
equation:
\be
\partial_t z = - \partial_x \left[
\partial_x^2 m + {m\over 1+m^2} - 
\lambda^* \partial_x \left( {1\over 1+m^2} \right)  \right]
+ \eta
\label{mod_asi}
\ee
for values of $\lambda^*$ varying between 0.1 and 1.

\begin{figure}
\begin{center}
\includegraphics[angle=-90,width=8.5cm,clip]{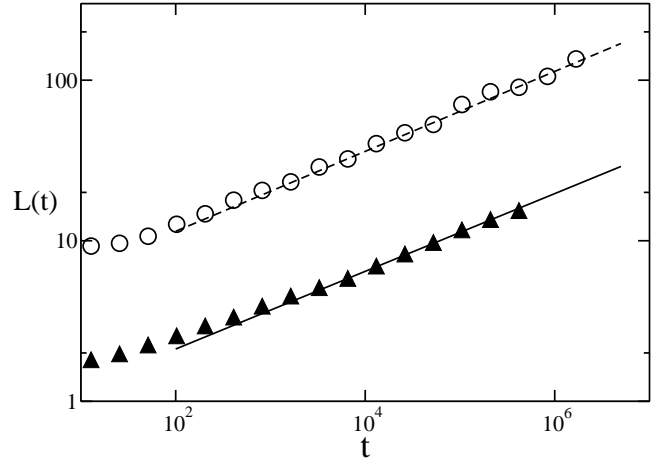}      
\end{center}
\caption{Coarsening for the asymmetric model 
(Eq.~(\protect\ref{mod_asi}) with $\lambda^*=1$),
in the absence (empty circles) and in the presence (full triangles)
of shot noise. Fits have been done for $t>10^4$ and give 
$n=0.25$ without noise (dashed line) and $n=0.24$ with noise (full line). 
Circles have been shifted by a constant. }
\label{fig:L_asi}
\end{figure}

Our results (see Fig.~\ref{fig:L_asi}) suggest that $\jsb$ is 
{\it irrelevant} for the
coarsening law, both for $\eta\equiv 0$ (deterministic case) and
for $\eta\ne 0$ (noisy case).

\section{Discussion and conclusions}
\label{sec:last}

\begin{figure}
\begin{center}
\includegraphics[angle=-90,width=8.5cm,clip]{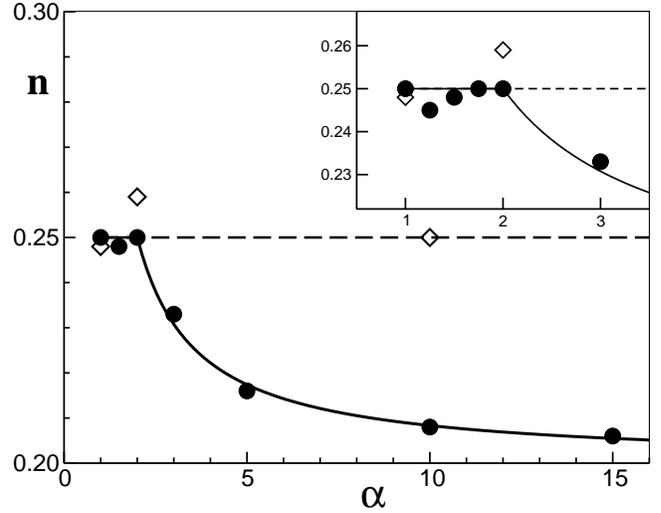}      
\end{center}
\caption{The coarsening exponent $n$ as a function of $\alpha$, for
the deterministic models (full circles) and for the stochastic models
(open diamonds). In the inset we enlarge the region of small $\alpha$.
Full line is the theoretical result in the absence of noise (Table~\ref{tab:n}) 
and the dashed line is our ansatz $n=1/4$ for the noisy case.}
\label{fig:n_alpha}
\end{figure}

In Fig.~\ref{fig:n_alpha} we have summarized our numerical and
theoretical results for the coarsening exponent $n$ ($L\sim t^n$).
In the absence of noise, our theory (full line)
predicts that $n=\fra{1}{4}$ for 
$1<\alpha\le 2$ and for larger values of $\alpha$ it decreases down to 
$\fra{1}{5}$ ($n=1/(5-\fra{2}{\alpha})$).
Numerical results (full circles) agree well with the full fine.

We are aware of  only 
one analytical paper treating our class of models (Ref.~\cite{Golub}).
The author uses scaling arguments to conclude that, {\it in the absence}
of noise, $n=1/4$ irrespectively of $\alpha$ and for any dimension $d$
of the substrate.
In the following we give a 
drastically simplified version of the scaling arguments.
If $Z,L$ and $M=Z/L$ are respectively the typical height, width and
slope of mounds at time $t$, the evolution equation 
for $z(x,t)$ implies $Z/t \sim j/L$. 
The current $j$ is made up of the Mullins term, of order $M/L^2$ plus the 
ES current, whose asymptotic expression for large slope is $1/M^{2\alpha -1}$. 
They vanish in the limit $t\to\infty$ and must be of the same order 
in $\fra{1}{t}$, which entails the relation 
$M^\alpha\sim L$. If their sum $j$ is supposed to be of the same order as 
well, the relation $Z/t \sim M/L^3$ implies $L(t)\sim t^{1/4}$, i.e.
$n=1/4$ for any $\alpha$.

One main drawback of the scaling argument
is that the two terms appearing
in $j$ are of the same order but their sum is smaller
(i.e. of higher order in $\fra{1}{t}$) because of a compensation effect.
This is a necessary condition for the stationary configurations
to play a role in the coarsening process.

In order to have a direct numerical check of our statement, we have
evaluated the quantities 
$\langle |\partial_x\jm |\rangle$,
$\langle |\partial_x\jes |\rangle$ and
$\langle |\partial_x(\jm + \jes) |\rangle$,
as a function of time,
where $\langle\cdots\rangle$ means, as before, the spatial average.
For all the considered values of $\alpha$ ($\alpha=1,3,10$),
the result is the same: the ratio 
$\langle |\partial_x\jm |\rangle/\langle |\partial_x\jes |\rangle$
is equal to one, up to higher order terms, and
$\langle |\partial_x(\jm + \jes) |\rangle/\langle |\partial_x\jm |\rangle$
vanishes.

Our numerical results tell even more than that: in fact scaling
arguments would suggest that $n$ is strictly smaller than $\fra{1}{4}$
if $j$ is smaller than $\jm$ and $\jes$, but for $\alpha\le 2$
we do find $n=\fra{1}{4}$ even if $|j|/|\jm|\simeq |j|/|\jes|\to 0$.

Model $\alpha=1$ had been previously studied numerically
at short times also in Ref.~\cite{Sander} and authors found a value 
$n\approx 0.22$, independent of the noise strength.

Let us now discuss the results in the presence of noise.
Fig.~\ref{fig:n_alpha} presents with diamonds the numerical results
for the stochastic integration of Eq.~(\ref{model}). Our data refer to 
$\alpha=1,2,10$ and provide a reasonably convincing evidence that
in the presence of noise the coarsening exponent remains constant,
$n=\fra{1}{4}$ (dashed line).
The somewhat larger value for $\alpha=2$ may be due to unknown
logarithmic corrections.

Authors in Refs.~\cite{noise1,noise2} use qualitative arguments to describe
coarsening assisted by noise: they use a `single mound' model and
find the coarsening time by requiring that shot noise
induces a height fluctuation of the same order of the mound height.
In one dimension they find $n=1/(3+\fra{2}{\alpha})$,
where $\alpha$ is defined phenomenologically through the asymptotic 
relation $M\sim L^{1/\alpha}$ between the typical (or the maximal)
slope $M$ and the width $L$ of mounds.

Their prediction for $n(\infty)$ seems to agree with the result 
$n=\fra{1}{3}$ for model~0. This is reasonable 
because in that model the slope is
constant and therefore it can effectively be equivalent to the model
$\alpha=\infty$.
Actually, if we take the limit $\alpha=\infty$ in 
Eq.~(\ref{2}), it is straightforward to
conclude that the current $\jes$ vanishes 
and we obtain the linear equation:
\be
\partial_t z(x,t) = -\partial_x^4 z(x,t) + \eta(x,t) ~.
\label{eq_lin}
\ee

The basic question is whether 
our class of $\alpha$-models tends --in some sense--
to Eq.~(\ref{eq_lin}) with increasing $\alpha$.
In the absence of noise, the answer is surely negative:
indeed, Eq.~(\ref{eq_lin}) does not admit stationary periodic 
solutions that, as discussed above, are crucial for
deterministic coarsening.

On the other hand,
if noise is present ($\eta\ne 0$), Eq.~(\ref{eq_lin}) describes a 
process of kinetic roughening: the growing surface is characterised
by a correlation length $\tilde\xi(t)\sim t^{1/\tilde z}$,
where $\tilde z$ is the dynamical critical exponent.
In $d=1$ it is well known~\cite{libroJV} that for the
quartic linear equation (\ref{eq_lin}), $\tilde z=4$.
It is reasonable to guess that 
the stochastic $\alpha$-model does converge to Eq.~(\ref{eq_lin}) 
for $\alpha\to\infty$ and --in the same limit-- $n(\alpha)\to 
\fra{1}{\tilde z}=\fra{1}{4}$.

The meaning of a constant value $n=\fra{1}{4}$ for any $\alpha$ 
is simple: in the presence of noise the detailed form of the
current $\jes$ is irrelevant provided that the slope $m$ diverges.

We can now summarize our main results.
Without noise, $n=\fra{1}{4}$ for $\alpha\le 2$ and 
$n=1/(5-\fra{2}{\alpha})$ for $\alpha>2$.
This result has been obtained analytically and it has been confirmed
by extensive numerical calculations.
It can not be deduced by simple scaling arguments.
In the presence of noise our numerical data for $\alpha=1,2,10$
suggest that $n=\fra{1}{4}$ irrespectively of $\alpha$.
This guess agrees with the well known result $\tilde\xi(t)\sim
t^{1/4}$, valid for the linear model $\alpha=\infty$.
So, steepening of mounds makes coarsening faster without noise
and slower with noise.

We believe that the surface profile can not be described
as a sequence of mounds with a spatially constant slope $M$ that
increases in time (see Fig.~\ref{fig:profili}, centre). 
This wrong assumption may be the reason why
qualitative arguments to determine $n$ do fail.

We have also considered the possible effect of a symmetry 
breaking term $\jsb$ in the current: 
it is irrelevant for model~1, as already proved for
model~0~\cite{kinks}.

We conclude the paper by mentioning a different model, whose
coarsening properties bear some similarities with our
$\alpha$-models. It has been
studied by Bray and Rutenberg~\cite{Bray} and it consists in 
the addition of a long-range attraction between kinks to model~0.
If such interaction decays as a power law of the distance 
($1/L^s$, with $s>1$)
the deterministic coarsening exponent is found to be $n=1/(1+s)$.
In the presence of conservative noise, this appears to be relevant for $s>2$
and in that case $n=\fra{1}{3}$.  Some analogies therefore exist 
with our class of $\alpha$-models, because in both cases
the coarsening exponent is a continuously varying
function of a parameter ($\alpha$ or $s$), noise may be relevant
($\alpha,s > 2$) or not ($1 < \alpha,s \le 2$), 
and --finally--
the stochastic coarsening exponent is constant if noise is relevant.

\begin{acknowledgement}
We warmly thank C. Castellano for a detailed and critical
reading of the manuscript. We also acknowledge useful discussions 
with A. Crisanti and D. Mukamel.
The readibility of the revised manuscript improved thanks to a careful
reading by S. Lepri and M. Moraldi.
\end{acknowledgement}

\begin{appendix}          

\section{Integration algorithms}

Let us rewrite in an explicit way the evolution equation
(\ref{model}) for the field $z(x,t)$:
\be
\partial_t z(x,t) = -\partial_x^4 z
-\partial_x
\left[\frac{\partial_x z}{(1+(\partial_xz)^2)^{\alpha}}
\right]
+ \eta(x,t) \quad ,
\label{pde}
\ee
where $\eta(x,t)$ indicates additive $\delta$-correlated 
spatio-tempo\-ral gaussian noise, i.e. 
\be
<\eta(x,t)> = 0
\ee
\be
<\eta(x,t) \eta(0,0)> = {\tilde F}_0 \delta(x) \delta(t)
\quad .
\ee

\subsection{Deterministic equation} 
\label{app:det}

Let us first neglect the noise term:
in order to perform the numerical integration of (\ref{pde})
we consider a discrete spatial grid of resolution 
$\delta x$ 
and a discrete time evolution with time step $\tau$.
The discretized field is written as
$z(i,n)$, where the integer indices $i$ and $n$ are
the spatial and temporal discrete variables, respectively.
Periodic boundary conditions have been
considered for the field: $z(i,n)=z(i+I,n)$,
where $I$ is the number of sites of the grid
($L_c=I \delta x$). The algorithm adopted to
integrate (\ref{pde}) is a time-splitting pseudo-spectral code
\cite{num}.  In particular, by following \cite{rdf}
Eq.(\ref{pde}) has been rewritten as
\be
\partial_t z(x,t) = ({\cal L} + {\cal N}) z(x,t)
\ee
where ${\cal L}$ and ${\cal N}$ are two operators
defined in the following way:
${\cal L} z =  -\partial_x^4 z$
and
${\cal N} z =  
-\partial_x\left[\frac{\partial_x z}{(1+(\partial_x z)^2)^{\alpha}}
\right]$. 
As usual for time splitting algorithms, the 
linear evolution, ruled by the operator
${\cal L}$, is treated independently
from the nonlinear one (associated to the operator ${\cal N}$).
A complete evolution over an integration time step $\tau$
therefore corresponds to the two successive integration steps: 
\be
 z^*(x,t+\tau) = \exp{[{\cal L} \tau]} z(x,t)
\ee
and
\be
 z(x,t+\tau) = \exp{[{\cal N} \tau]} z^*(x,t+\tau)
\ee
where $z^*(x,t)$ is a dummy field.

Let us firstly consider the linear part,
\be
\partial_t z(x,t) = {\cal L} z(x,t) ~.
\label{lin}
\ee

Eq.(\ref{lin}) can be easily solved in the 
Fourier space and the equation of motion for the 
spatial Fourier transform of the field ${\tilde z}(p,t)$ is
\be
\partial_t {\tilde z}(p,t) = -p^4 {\tilde z}
\quad .
\label{fftlin}
\ee
The time evolution for ${\tilde z}$ is simply given by
\be
{\tilde z}^* (p,t+\tau) = \exp{[-p^4\tau]} {\tilde z} (p,t) ~ .
\label{linsol}
\ee
Therefore, in order to integrate Eq.(\ref{lin}), 
the field should be Fourier transformed in space (${\cal F}$), 
then multiplied by the propagator reported in Eq.(\ref{linsol})
and the outcome of such operation should be finally inverse-Fourier
transformed (${\cal F}^{-1}$):
\be
z^* (x,t+\tau) = {\cal F}^{-1} \exp{[-p^4\tau]} 
{\cal F} z (x,t) \,.
\label{linsol2}
\ee

The integration of the nonlinear part has been performed
by emploing a second-order Adam-Basforth scheme
\bea
z(x,t+\tau) &=&  z^*(x,t+\tau) + \\
& &\frac{\tau}{2} \left[ 3 \, G(z(x,t))
- G(z(x,t-\tau)) \right] \quad ,
\nonumber
\label{solnonlin}
\eea
where $G(z(x,t)) =
\left\{-\partial_x\left[\frac{\partial_x z}
{(1+(\partial_x z)^2)^{\alpha}}
\right]\right\} $.
In order to obtain a better precision, the spatial 
derivatives appearing in $G(z(x,t))$ have been evaluated 
in the Fourier space.

\subsection{Stochastic equation} 
\label{app:noise}

Let us now consider the noisy problem: in this case
the algorithm outlined here above does not guarantee a
sufficient precision. Therefore, we have developed
a more accurate integration scheme \cite{cris}
that consists as a first step in rewriting the equation of motion
(\ref{pde}) in the Fourier space
\be
\partial_t {\tilde z}(p,t) = -p^4 {\tilde z}(p,t)
+G_p(z,t) + {\tilde \eta} (p,t)\, ,
\label{pde_fft}
\ee
where $G_p(z,t)$ and ${\tilde \eta} (p,t)$
are the Fourier transforms of the nonlinear
part and of the noise term appearing in Eq.~(\ref{pde}). 
The amplitudes ${\tilde \eta} (p,t)$ of the noise components 
in the Fourier space are still gaussian $\delta$-correlated
stochastic variables with zero average and
with a variance ${\cal V}_p = {\cal F}_0/ I$ independent of $p$ (white noise).
The formal exact solution of (\ref{pde_fft}) is
\bea
&&{\tilde z}(p,t+\tau) = \\
&&{\rm e}^{-p^4\tau} 
\left[ {\tilde z}(p,t) +\int_t^{t+\tau} dt' \, {\rm e}^{p^4(t'-t)} 
(G_p (z,t') + {\tilde \eta} (p,t'))\right] ~ .
\nonumber
\label{int1}
\eea

The problem is now to evaluate the two terms appearing
in the integral.
The first term has been evaluated adopting a second order
Adam-Bashfort scheme. i.e.
\bea
&&\int_t^{t+\tau} dt' \, {\rm e}^{p^4(t'-t)} 
G_p (z,t') = \\
&&\frac{\tau}{2} \left[ 3 G_p(z,t) -
{\rm e}^{-p^4\tau} G_p(z,t-\tau) \right]
+ {\cal O}(\tau^3) \,.
\label{int2}
\nonumber
\eea

The treatment of the second term is more delicate, since
it is a stochastic integral:
we have chosen to evaluate it accordingly
to the Ito's prescription~\cite{gardiner}:
\bea
&&\int_t^{t+\tau} dt' \, {\rm e}^{p^4(t'-t)} 
{\tilde \eta} (p,t') = \\
&&\left[ W_p (t+\tau) - W_p (t) \right] = \Delta W_p(t)
\,.
\nonumber
\label{int3}
\eea

Here $W_p(t)$ and $\Delta W_p(t)$ represent two Wiener processes:
in particular we have $\langle\Delta W_p\rangle= 0$ and 
$\langle(\Delta W_p)^2\rangle = \tau {\cal V}_p$.
The complete solution of (\ref{pde_fft}) can be written as
\bea
&&{\tilde z}(p,t+\tau) = {\rm e}^{-p^4\tau}
{\tilde z}(p,t) + \\
&&\frac{3 \tau}{2} {\rm e}^{-p^4 \tau} G_p (z,t) 
-\frac{\tau}{2} {\rm e}^{- 2 p^4 \tau} G_p (z,t-\tau) + 
{\rm e}^{-p^4 \tau}
\Delta W_p (t) \,.
\label{sol}
\nonumber
\eea
In order to obtain the solution in the real space it is
sufficient to inverse-Fourier transform ${\tilde z}(p,t+\tau)$.

Due to the spatial and temporal discreteness of the
integration scheme, the spatio-temporal noise term 
$\eta (x,t)$ should be rewritten as $\gamma_i^n$, 
where $\gamma$ is a random gaussian
variable of zero average and variance
\be
{\cal V}=\frac{\tau \tilde F_0}{\delta x} ~ ,
\ee
with
\be
<\gamma_i^n \gamma_0^0 > = {\cal V} \delta_{i,0} \delta_{n,0}~.
\ee

The spatio-temporal discrete gaussian noise, with zero average 
and standard deviation $\sqrt{\frac{\tau \tilde F_0}{\delta x}}$
has been numerically generated by
employing a Box-Muller algorithm \cite{num}.

\end{appendix}


\begin{thebibliography}{99}
\bibitem{note2}
A linear stability analysis of a 1d vicinal surface of constant 
slope $m^*$ shows that step-flow is stable if $\jes'(m^*)<0$.
\bibitem{nota1}
Strictly speaking negative eigenvalues are a necessary but not a
sufficient condition to ensure coarsening.
\bibitem{nota2}
The exact relation, as derived by Langer~\protect\cite{Langer}
is $t^{-1}=\sum_q |\epsilon(q,L)|$ where the sum is restricted 
to the negative (and therefore unstable) modes. However, the $L$
dependence of $\epsilon$ can be pulled out the summation and it is
therefore sufficient to consider the ground state energy.
\bibitem{nota3} 
Joining means that $\Ud(x)=\Ui(x)$ for $x\le\fra{L}{2}$ and
$\Ud(x)=\Ui(x-L)$ for $x\ge\fra{L}{2}$.
This kind of juxtaposition is used by Landau and Lifshitz~\protect\cite{LL}
in their treatment of the double well problem and for model~0 it gives the
correct results.
\bibitem{nota4}
The Newton's equation for the model $\alpha=3$ can be easily integrated 
if the particle has zero energy and therefore an
explicit expression for $\Ui(x)$ is available. We do not report it here,
but we write explicitly the first correction to the leading order in the
large $x$ expansion: $\Ui(x)\simeq \fra{10}{9x^2} + 
(\fra{16}{9})^{1/3}\fra{1}{x^{8/3}}$. At $x=10$ the first correction is
of order of one quarter of the leading term.
\bibitem{nota_fig}
In order to find the ground state energy
we have imposed periodic boundary conditions and searched
for the zero node wavefunction. Precision real$\ast$16 has been
necessary in order to achieve a sufficient accuracy at large $L$.
\end{thebibliography}
\end{document}